\documentclass[12pt,preprint]{aastex}
\def\omega0{\Omega_{\rm m,0}}

\def\rv{r_{\rm vir}}
\def\rs{r_{\rm s}}

\def\kms{\,\rm km\,{s}^{-1}}
\def\kpc{\,\rm kpc}
\def\mpc{\,\rm Mpc}
\def\LCDM{\Lambda{\rm CDM}}
\def\Ds{D_{\rm s}}
\def\sigmasim{\sigma_{\rm 1''}}
\def\tausim{\tau_{\rm 1''}}

\def\deff{d_{\rm eff}}
\def\msun{M_\odot}

\def\ns{n_{\rm s}}

\def\zs{z_{\rm s}}
\def\zl{z_{\rm l}}
\def\beq{\begin{equation}}
\def\eeq{\end{equation}}
\def\xcenter{x_{\rm center}}
\def\pppm{\,\rm P^3M}
\begin{document}

\title{Is the Number of Giant Arcs in $\LCDM$ Consistent With Observations?}
\author{Guo-Liang Li\altaffilmark{1,4}, S. Mao\altaffilmark{2}, Y.P. Jing\altaffilmark{1},
 M. Bartelmann\altaffilmark{3}, X. Kang\altaffilmark{1}, M. Meneghetti\altaffilmark{3}
}
\altaffiltext{1}{Shanghai Astronomical Observatory; the Partner Group
of MPA, Nandan Road 80, Shanghai 200030, China; Email:
{\tt (lgl,ypjing, kangx)@center.shao.ac.cn}}
\altaffiltext{2}{ University of Manchester, Jodrell Bank Observatory,
  Macclesfield, Cheshire SK11 9DL, UK; Email: {\tt smao@jb.man.ac.uk}}
\altaffiltext{3}{Zentrum f\"ur Astronomie, ITA, Universit\"at
  Heidelberg, Albert-\"Uberle-Str. 2, 69120 Heidelberg, Germany; Email:
{\tt (mbartelmann, meneghetti)@ita.uni-heidelberg.de}
}
\altaffiltext{4}{Graduate School of Chinese
Academy of Sciences, Beijing 100039, China}

\shorttitle{Giant Arcs in $\LCDM$}
\shortauthors{Li et al.}

\begin{abstract}
We use high-resolution N-body simulations 
to study the galaxy-cluster cross-sections and the abundance of 
giant arcs in the $\Lambda$CDM model. Clusters are selected from the
simulations using the friends-of-friends method,
and their cross-sections for forming giant arcs are
analyzed. The background sources are assumed to follow
a uniform ellipticity distribution from 0 to 0.5 and to have
an area identical to a circular source with diameter $1\arcsec$.  
We find that the optical depth scales as the source redshift
approximately as 
$\tau_{1''} = 2.25 \times 10^{-6}/[1+(\zs/3.14)^{-3.42}]$
($0.6<\zs<7$). The amplitude is about 50\% higher
for an effective source diameter of $0.5\arcsec$.
The optimal lens redshift for giant arcs with the
length-to-width ratio ($L/W$) larger than 10 increases from 0.3 for $\zs=1$,
to 0.5 for $\zs=2$, and to 0.7-0.8 for $\zs>3$. 
The optical depth is 
sensitive to the source redshift, in qualitative agreement with Wambsganss et
al. (2004). However, our overall optical depth
appears to be only $\sim$ 10\% to 70\% of those from
previous studies. The differences can be mostly explained by different power
spectrum normalizations ($\sigma_8$) used and different ways
of determining the $L/W$ ratio. Finite source size
and ellipticity have modest effects on the optical depth.
We also found that the number of highly magnified (with magnification
$|\mu|>10$) and ``undistorted'' images (with $L/W<3$)
is comparable to the number of giant arcs with $|\mu|>10$ and
$L/W>10$.
We conclude that our predicted rate of giant arcs may be 
lower than the observed rate, although the precise `discrepancy' is still
unclear due to uncertainties both in theory and observations.
\end{abstract}

\section{INTRODUCTION}

The Cold Dark Matter (CDM) scenario is
now the standard model for structure formation. The
``concordance'' $\LCDM$ model (e.g., Ostriker et al. 1995)
is supported by many observations, in particular the
large-scale structure of the universe (e.g. Jenkins et al. 1998;
Peacock et al. 2001; Tegmark et al. 2004) and the cosmic microwave
background (e.g., Spergel et al. 2003).

Giant arcs are formed when background galaxies are distorted into long
arc-like shapes
by the gravitational shear of intervening clusters of galaxies. They are among the most
beautiful images on the sky (e.g., see the Hubble Space Telescope [HST] images
of A2218, Kneib et al. 1996). As clusters of galaxies
are the largest bound structures in the universe, they are at the
tail of the mass function of bound structures. The formation of giant arcs
therefore critically depends on the abundance and density profiles of clusters.
Observationally, the number of giant arcs was first estimated using the Einstein
Medium-Sensitivity Survey (EMSS) by Gioia \& Luppino (1994).
The largest dedicated search for giant arcs in X-ray clusters was performed by
Luppino et al.
(1999) who found strong lensing in 8 out of 38 clusters. 
These fractions were confirmed by
Zaritsky \& Gonzalez (2003) using the Las Campanas Distant Cluster
Survey and Gladders et al. (2003) using the Red-Sequence Cluster Survey. 
A recent extensive analysis of HST
images found 104 candidate tangential arcs in 128 clusters with
length to width ratios exceeding 7 (Sand et al. 2005)\footnote{Their
  $L/W$ ratio is similarly defined as in Dalal et al. (2004) where giant
arcs are fitted with a rectangle rather than an ellipse, so their
$L/W$ ratio is $4/\pi$ our value.}.
All
these recent studies show that giant arcs are quite common in clusters
of galaxies.

Earlier predictions of giant arcs use simple spherical models (e.g., Wu
\& Hammer 1993). However, such models are clearly
inadequate, as the ellipticities and substructures of clusters can enhance the lensing
cross-section by a large factor (Bartelmann \& Weiss 1994;
Bartelmann, Steinmetz \& Weiss 1995). More recently, Torri et al. (2004)
studied the importance of mergers for arc statistics. Wu \& Mao (1996)
compared the arc statistics in the Einstein-de Sitter universe and
$\Lambda$CDM using simple spherical cluster
models and found a factor of $\sim 2$ increase in the $\Lambda$CDM
model. A much more realistic study was performed by
Bartelmann et al. (1998, hereafter B98) who first pointed out that the number of giant
arcs in the $\LCDM$ model appears to be below the observed
rate in clusters by a factor of 5-10. Meneghetti et al. (2000)
investigated the effects of cluster galaxies on arc statistics and found that cluster galaxies 
do not introduce perturbations strong enough to significantly change 
the number of giant arcs. Flores, Maller \& Primack (2000) drew a similar
conclusion. Meneghetti et al. (2003) further examined the
effect of cD galaxies on the giant arcs and concluded they are
insufficient to resolve the discrepancy. Oguri et al. (2003) studied the
ability of tri-axial models of dark matter halos to form giant arcs. They
concluded that an inner power-law 
profile of $r^{-1.5}$ can reproduce the observed
giant arcs, while the standard NFW profile (Navarro, Frenk, \& White
1997) with an inner profile of $r^{-1}$ is unable to match the
observations. Bartelmann et al. (2003) and Meneghetti et al. (2005) studied 
the probability of the formation of giant arcs in galaxy clusters in different 
dark-energy cosmologies. But the effect is still insufficient.
In an important paper, Wambsganss et al. (2004) pointed
out the simple fact that the probability of high lensing magnifications
is very sensitive to
the assumed source redshift. For example, a source at $\zs=1.5$ has a factor
of $\sim 5$ higher optical depth than a source at redshift $\zs=1$.
As the redshift distribution for the background source population that
forms arcs is not well known, this introduces an uncertainty in the
comparison between observations and predictions.

In this paper, we use high-resolution simulations to re-examine
the predicted number of giant arcs and compare our predictions
with observations. The outline of the paper is as follows. In \S2,
we discuss the numerical simulations we use and the method we
employ to identify giant arcs in simulated clusters. In \S 3, we
present the main results of the paper and compare them
with several previous studies (Bartelmann et al. 1998;
Wambsganss et al. 2004; Dalal et al. 2004). In particular, we confirm
the results of Wambsganss et al. (2004; see also Dalal et al. 2004)
that the predicted arc number is sensitive to the assumed source redshift distribution.
However, our overall optical depth appears to be lower than these
previous studies, sometimes by a large factor. We examine in detail how the
differences arise. This reduction makes it somewhat difficult
to match the observed high frequency of giant arcs with our predictions.

In \S4, we summarize our results and discuss areas for future improvements.

\section{Methods}

\subsection{Numerical Simulations}

The cosmological model considered here is the current popular
$\Lambda$CDM model with the matter-density parameter $\omega0=0.3$ and the
cosmological constant $\Omega_{\Lambda,0}=0.7$. The shape parameter
$\Gamma=\omega0 h$ and the amplitude $\sigma_8$ of the linear density
power spectrum are taken to be 0.21 and 0.9, respectively, where
$h$ is the Hubble constant in units of $100\kms\mpc^{-1}$ and we take $h=0.7$.
A cosmological N-body
simulation with a box size $L=300h^{-1} \mpc$, which was generated with our
vectorized-parallel $\pppm$ code (Jing \& Suto 2002; Jing 2002), is used in
this paper. The simulation uses $512^3$ particles, so the particle
mass $m_{\rm p}$ is $1.67\times 10^{10}h^{-1}\msun$. The gravitational force is
softened with the $S2$ form (Hockney \& Eastwood 1981) with the
softening parameter $\eta$ taken to be $30h^{-1}\kpc$. The simulation has similar mass and
force resolutions as that used by B98, but our
simulation volume is 8 times larger, and hence contains 
$\sim 8$ times more clusters. Compared with the simulations of
Wambsganss et al. (2004, hereafter W04),
our resolutions are a factor of $6.6$ lower in mass and a
factor of 10 lower in the force softening. Since strong arcs are produced
mostly at radii $\ga 100h^{-1} \kpc$ in the lens plane, the resolutions
are sufficient (see also Dalal et al. 2004). This is also supported
by the comparison of our results with those of W04 which
are based on much higher resolution -- our predicted number of
giant arcs is in reasonable agreement with that of W04 if the
arcs are identified with their method (see \S\ref{sec:comparisonsW04} and 
Fig. \ref{fig:tau}).

Dark matter halos are identified with the friends-of-friends 
method using a linking length equal to 0.2 times the mean
particle separation. The halo mass $M$ is defined as the virial mass
enclosed within the virial radius according to the spherical collapse
model (Kitayama \& Suto 1996; Bryan \& Norman 1998; Jing \& Suto 2002).

\subsection{Lensing Notations}

For convenience of later discussions, we outline the notations we will
use in the paper. We denote the lensing potential as $\phi$. 
The mapping from the lens and source plane is given by
\beq
\overrightarrow{y} = \overrightarrow{x}-\nabla \phi  ,
\eeq
and the distortion of images is then described by the Jacobian
\beq
{\partial y_i
 \over \partial x_j}
=
 \left( \begin{array}{cc} 1-\phi_{11} & -\phi_{12} \\
-\phi_{21} & 1-\phi_{22} \\
\end{array} \right),
\eeq
where $(y_1, y_2)$ are dimension-less source coordinates in the source plane, and
$(x_1, x_2)$ are dimension-less image coordinates in the lens plane.
The Jacobian matrix is given in terms of the lensing potential by the
$2\times2$ matrix on the
right hand side, which involves second-order derivatives of $\phi$
with respect to
$x_1$ and $x_2$. The eigenvalues of the Jacobian matrix
are denoted as $\lambda_1$ and $\lambda_2$. Without losing generality,
we will assume $|\lambda_1| \ge |\lambda_2|$. The
(signed) magnification is given by
\beq
\mu = {1 \over \lambda_1 \lambda_2}.
\eeq
For an infinitesimal circular source, the length to width ratio 
is simply given by
\beq
{L \over W} = {{1/|\lambda_2|} \over {1/|\lambda_1|}}  
=\lambda_1^2 |\mu|.
\label{eq:lw-mu}
\eeq
Notice that $L/W$ is not equal to $|\mu|$ if $\lambda_1$ is not unity.

\subsection{Lensing Simulations}

The lensing properties of numerical clusters are studied using the
ray-tracing technique (e.g., B98; W04). For the source population, we use redshifts
ranging from 0.6 to 7 with a uniform interval of 0.4. To calculate
the optical depth,  we use 20 outputs for the simulated box from
redshift 0.1 to 2.35 with a step size of $\sim 0.1$. For each redshift, we
select the 200 most massive clusters of galaxies as lenses. Notice
that this cluster sample is more than a factor ten larger than the
samples used by Bartelmann et al. (1998) and Ho \& White 
(2004), although our redshift
samplings are sparser than the most recent study of 
Torri et al. (2004). For each cluster, we use the particles within
a cube with a side-length of two virial radii ($2\rv$). The surface densities are
then calculated for three orthogonal projections using the SPH smoothing
algorithm (Monaghan 1992) on a $1024\times 1024$ grid. Usually, the kernel size is taken to be $30h^{-1}\kpc$
  (comoving). If the particle number within the kernel is fewer than 64,
 then we double the
  kernel size until the particle number is larger than 64. However, for
  the high density regions, when the particle number is larger than 400
  within the kernel, we only use the nearest 400 particles
  to estimate the density. In
order to obtain the lensing potential, we also use a 
larger grid ($2048\times 2048$) centered on the smaller grid; the
surface densities are padded with zeros for all the pixels outside the
inner grid. The lensing potential is then obtained using the FFT method
(B98). Notice that the larger grid is used to avoid aliasing problems due to periodic
boundary conditions.

To perform efficient lensing simulations, we first identify
regions of interest in the lens plane that have magnifications
exceeding $\mu_{\rm limit} \equiv 2.5$ and regions with $\mu\rm < 0$; 
other regions are unlikely to produce giant arcs
with $L/W$ exceeding 7.5. Following B98, we assume the
sources have an ellipticity, which is equal to one minus the axis ratio,
randomly distributed between 0 to 0.5.
Each source has an area $S_{\rm source}$ equal to a circular source
with a diameter of $1\arcsec$. To have sufficient numbers of pixels in one
image, the resolution in the regions of interest is increased to $0.1\arcsec$ (see
below) so that most images have at least $\ns=\pi \times
0.5^2 \times \mu_{\rm limit} /0.1^2 \approx 200$ pixels.

For the regions identified above, the
$1024\times 1024$ (coarse) grid does not provide
sufficient resolution. To remedy this, we
obtain the lensing potential on a finer grid with
resolution of $0.1\arcsec$ using cubic spline interpolation of the 
surrounding $14\times 14$ coarse grid points. In the same step, we obtain
the source position for each grid point,
and the corresponding magnification and eigenvalues 
($\lambda_1$ and $\lambda_2$).

Once we obtain the lens mapping for the regions of interest
from the lens plane to the source plane,
we locate the smallest rectangle with area $S_{\rm box}$ that contains
all the mapped pixels in the source plane. We then put sources randomly in this
rectangle and obtain their imaging properties. The mapped regions
in the source plane usually have irregular shapes, so the rectangle
will enclose not only the regions of
interest but also some regions that do not satisfy our selection
criteria ($\mu\ge \mu_{\rm limit} \equiv 2.5$ or $\mu < 0$). 
To avoid sources that straddle the irregular boundary,
we only analyze sources further if they contain at least
$\ns (\approx 200)$ pixels. In order to sample the
regions of arc formation well, a large number of sources are placed
randomly inside the regions of interest.  The number we generate is given by
$n_{\rm source} =9 {S_{\rm box} /S_{\rm source}} 
$, where $S_{\rm source}=\pi \times 0.5^2$ in square arc seconds,
and usually we have $n_{\rm source} > 10^4$.

For the pixels contained in a source inside the region of interest,
we use the lens mapping already derived to obtain the
corresponding image positions, and then use the friends-of-friends
method to identify giant arcs. To obtain the $L/W$ ratio, we assume 
giant arcs can be approximated as ellipses, hereafter we refer to this
method as ellipse-fitting. Numerically we find
this assumption to be valid for most of our giant arcs. However,
some images are irregular and cannot be fitted well by an ellipse. They
occur when the source straddles higher-order singularities (such as
beak-to-beak) or when the source size is comparable to the
caustics. But such cases are rare ($\sim 1\%$) for clusters that
contribute most of the lensing cross-sections  and 
will thus not affect our results significantly. To calculate
the $L/W$ ratio, we need to identify the center of a giant arc,
$\xcenter$, which is first taken to be the center of light. Due to
the (curved) arc shape, the center often falls outside the arc itself.
To find a better center,
we select one half of the image positions that are
closest to the current center and then re-calculate the center of light.
The process is repeated until the center does not change significantly.
We verified visually that this procedure yields reliable centers of
giant arcs. Once the center is identified, we then locate
the point $x_1$ that is the furthest from
$\xcenter$, and finally the
point $x_2$ that is the furthest from $x_1$. We 
fit a circular arc that passes through these three points. The length of
the arc is taken to be twice the major axis length of the ellipse, $a$, and
the length of the minor axis is taken to be $b=S_{\rm image}/(\pi
a)$, where $S_{\rm image}$ is the area covered by the image. The
$L/W$ ratio is then simply $a/b$. This procedure is
identical to the ellipse fitting method in B98.

In W04, the $L/W$ ratio is approximated by the
magnification. In order to obtain the arc cross-sections, 
the magnification patterns need to be calculated on a grid in the
source plane. In W04, each pixel has a size of $1.5\arcsec$.
However, we find that this is too large for small clusters in our simulations,
so we instead used a grid of $0.1\arcsec$ for all the clusters.
We use the Newton-Raphson method to find the
image position and its corresponding magnification and
eigenvalues. When there are 
multiple images, following W04, we use only the image with the largest
(absolute) magnification. In the following, we will study the validity of 
two approximate measures of $L/W$, namely the magnification, $|\mu|$, and
the ratio of the two eigenvalues, $|\lambda_1/\lambda_2|$. 
Under these two approximations, we obtain two corresponding
cross-sections, which we will denote as
$\sigma_\mu$ and $\sigma_\lambda$. We
will compare these two cross-sections with that obtained through
ellipse-fitting, $\sigmasim$. The corresponding 
optical depths can be obtained by integrating the cross-sections along
the line of sight (cf. Eq. \ref{eq:tau}), and we will denote these
as $\tausim$, $\tau_{\lambda}$ and $\tau_{\mu}$. The subscript $1\arcsec$
indicates that the sources have an effective 
diameter of $d_{\rm eff}=1\arcsec$. Later we will study lensing
cross-sections and optical depths for other effective source diameters,
$d_{\rm eff}=0.25\arcsec, 0.5\arcsec, 1.5\arcsec$, and their cross-sections
and optical depths will be labeled accordingly.

\section{Results}

\subsection{Caustics, magnifications and cross-sections
  \label{sec:caustics}}

\begin{figure}
\epsscale{0.8} 
\plotone{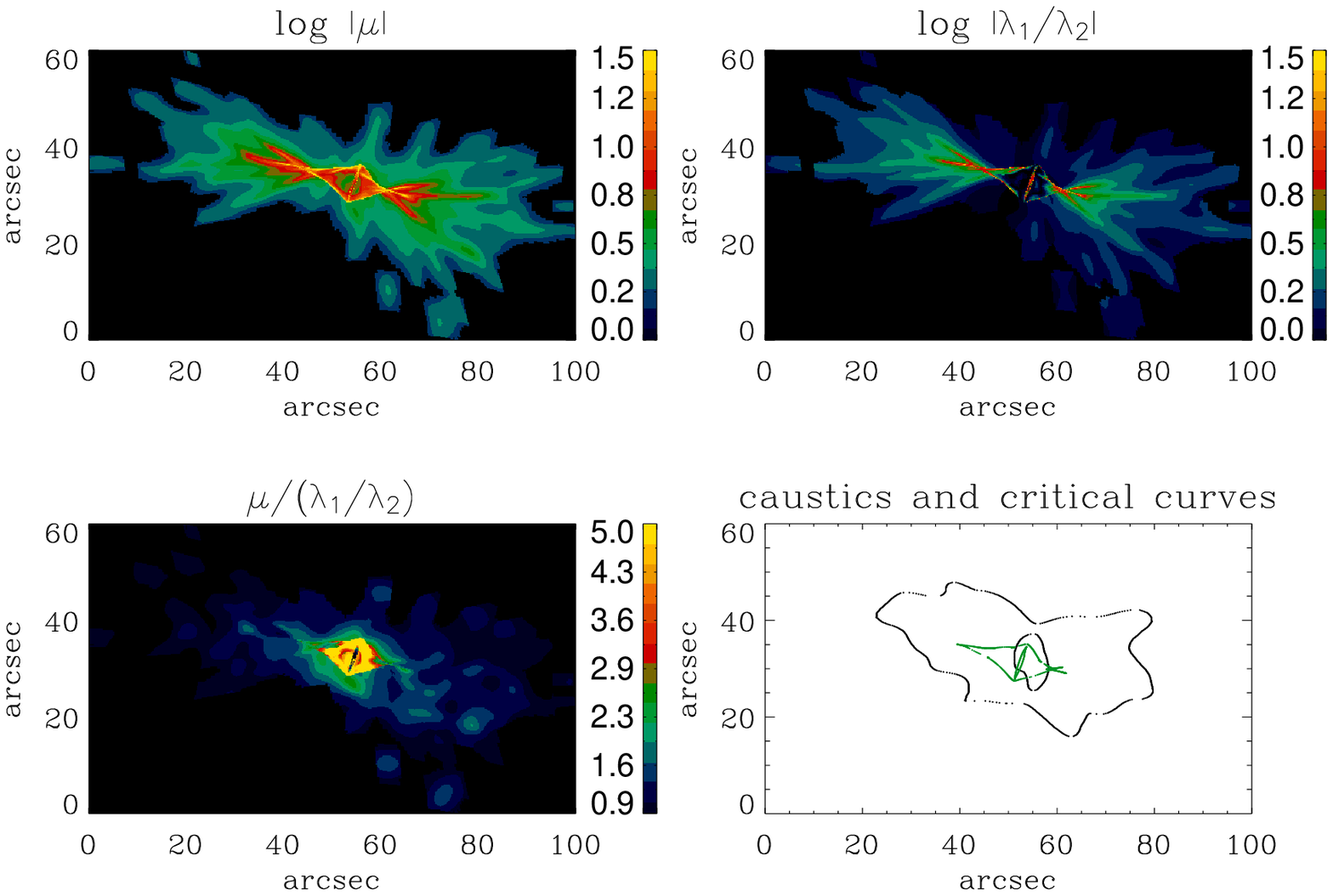} 
\caption{Maps of the absolute magnification (top left), ratio of the two
eigenvalues
$|\lambda_1/\lambda_2|$ (top right), $\mu/(\lambda_1/\lambda_2)$
(bottom left), and
the caustics and critical curves (bottom-left panel) are shown for the
fifth most massive cluster  ($M=7.58 \times 10^{14}h^{-1} M_\odot$) in
our simulations at redshift 0.3. The
sources are assumed to be at redshift 1.0.
The $|\mu|$ and $\lambda_1/\lambda_2$ values
in regions where $0<\mu<2.5$ are set to unity. Pixels with
values exceeding the maximum of the color bars 
are set equal to the maximum.
}
\label{fig:4p}
\end{figure}

\begin{figure}
\epsscale{0.8} \plotone{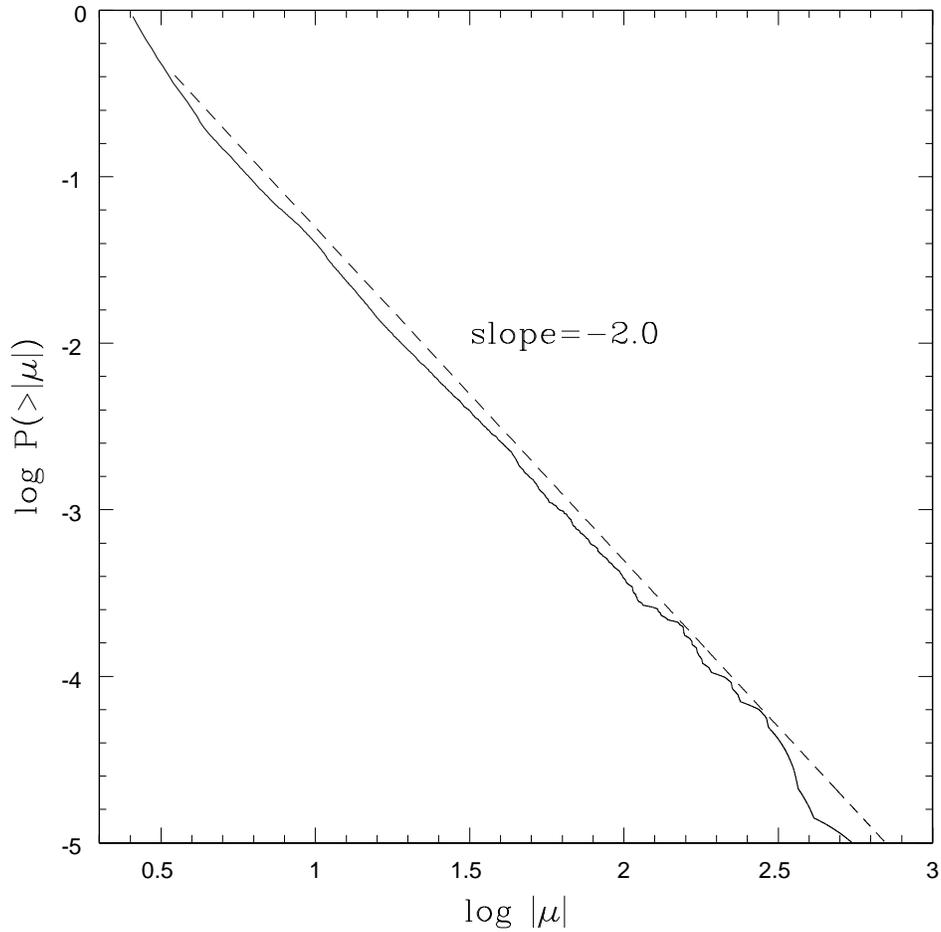} 
\caption{ The fraction of area
where the magnification is larger than a given $|\mu|$ in the
source plane. The data are taken from one realization of
the projected surface density for the fifth most massive cluster
(shown in
Fig. \ref{fig:4p}). The behavior can  be well fitted by the expected
asymptotic power-law,
$P(>|\mu|)\propto\mu^{-2} $, as indicated by the dashed line.
}
\label{fig:mu}
\end{figure}

For illustrative purposes, 
Fig. \ref{fig:4p} shows the lensing properties for 
the fifth most massive cluster ($M=7.58 \times 10^{14} h^{-1} M_\odot$)
in our simulation at redshift 0.3.
The source is at redshift 1. The
bottom right panel shows the critical curves and the caustics.
The other three panels show the maps of  $|\mu|$, $|\lambda_1/\lambda_2|$ and
$\mu/(\lambda_1/\lambda_2)=1/\lambda_1^2$
in the source plane. Recall that we only consider 
regions where $\mu>\mu_{\rm limit} \equiv 2.5$ or $\mu<0$. In other regions, we set
$\mu$ and $\lambda_1/\lambda_2$ to unity, which are
shown as  black in the top left panel.
The color bars show the range of the quantities 
plotted in the maps. Pixels
with values exceeding the maximum of the color bar are set equal to
the maximum. If we approximate the $L/W$ ratio
as either $|\mu|$ or $|\lambda_1/\lambda_2|$, the top two
panels clearly indicate that the cross-section
$\sigma_\mu$ will be much larger than the cross-section $\sigma_\lambda$
for a given length-to-width ratio. This effect is illustrated further in
the bottom left panel where we plot the map of $\mu/(\lambda_1/\lambda_2)$.
In most cases, especially for the high magnification regions
where arcs are expected to form, this ratio is larger than 1, which
indicates that for an infinitesimal circular source, $|\mu|$ will
over-estimate the $L/W$ ratio.
As a check of our lensing simulations, Fig. \ref{fig:mu} shows the
magnification probability distribution for the cluster shown in Fig. \ref{fig:4p}.
Our results nicely reproduce the asymptotic relation $p(>|\mu|)
\propto \mu^{-2}$ expected from the fold caustics when $|\mu| \gg
1$ (e.g., Schneider, Ehlers, \& Falco 1992).

\begin{figure}
\epsscale{0.8} \plotone{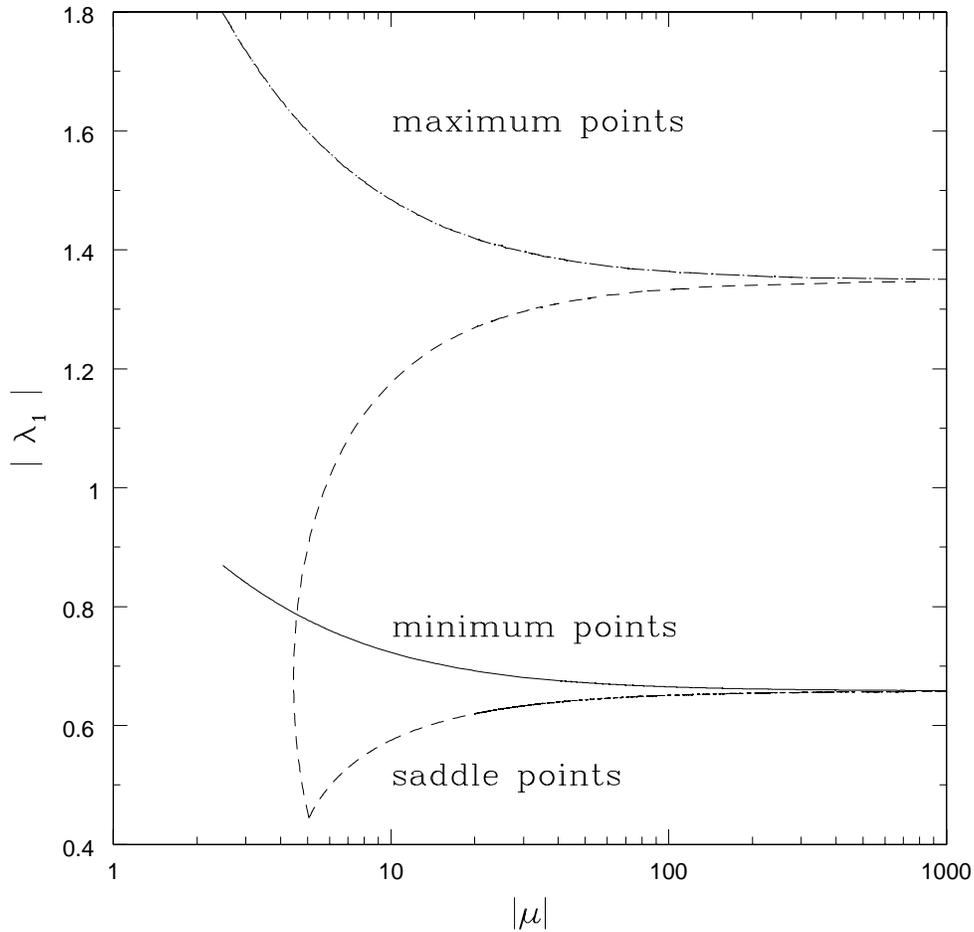} 
\caption{Magnification $|\mu|$ vs. the larger of the two eigenvalues for
a cluster described by a generalized NFW model (eq. \ref{eq:gnfw}) with
the power-law index
$\alpha=1.5$, concentration parameter $c=2.28$, and a total mass of
$M=10^{15} h^{-1} M_{\odot}$. The solid, dashed, and dash-dotted lines
are for the minimum, saddle and maximum images in
the time delay surface respectively. For an infinitesimal circular source,
$\sigma_\mu/\sigma_\lambda \approx \lambda_1^4$ for large magnifications
(see the text).
}
\label{fig:gnfw}
\end{figure}

To understand why the approximation $L/W=|\mu|$ appears to over-estimate
the arc cross-section, we study a cluster with a generalized
NFW profile (Navarro, Frenk \& White 1997; see also Moore et al. 1998):
\beq
\rho(r)\propto r^{-\alpha}(r+\rs)^{-3+\alpha}, ~~ 0<r<\rv,
\label{eq:gnfw}
\eeq
where $\rs$ is the scale radius, and $\rv$ is the virial radius.
The cluster is at redshift of 0.3
with a mass of $10^{15} h^{-1} M_{\odot}$. The power-law index ($\alpha$) is taken
to be 1.5 and the concentration parameter $c=\rv/\rs=2.28$ (Oguri 2002).
The source redshift is taken to be unity.
Fig. \ref{fig:gnfw} shows the relation between $|\lambda_1|$ and $|\mu|$
for the minimum, saddle and maximum images in the time delay surface.
The tangential arcs are primarily formed by the minimum images while
the radial arcs are formed by the maximum images.
For an infinitesimal circular source, the $L/W$ ratio is equal to $|\mu|$
multiplied by $\lambda_1^2$ (see Eq. \ref{eq:lw-mu}). It is clear
that for the minimum image, at high
magnifications, $\lambda_1 \rightarrow 0.65$, while for the maximum
image, $\lambda_1 \rightarrow 1.3$. As 
for high magnifications, the
asymptotic cross-section follows the probability distribution $p(>|\mu|)
\propto \mu^{-2}$ (Schneider et al. 1992), the cross-sections
for the tangential arcs therefore satisfy
$\sigma_\lambda \approx {\lambda_1}^4\sigma_\mu 
\approx 0.18 \sigma_\mu$. Similarly, for the radial arcs, we have 
$\sigma_\lambda \approx {\lambda_1}^4\sigma_\mu 
\approx 2.9 \sigma_\mu$. As the tangential arc cross-section usually
dominates over the radial arc cross-section, it follows that
the  $L/W=|\mu|$ assumption will over-estimate the
arc cross-sections by a factor of $\approx 1/0.18=5.6$.

\begin{figure}
\epsscale{1.1} \plottwo{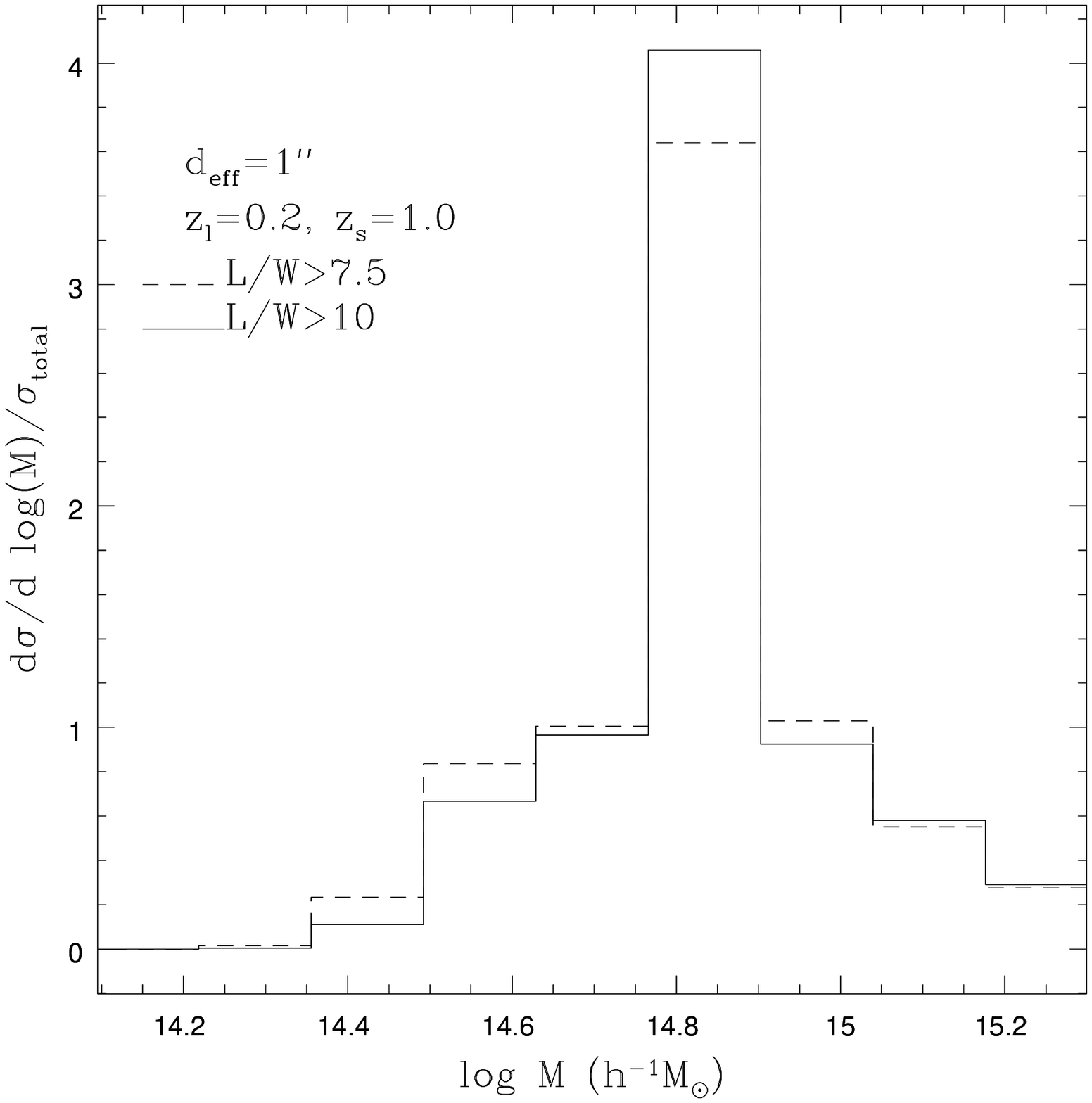}{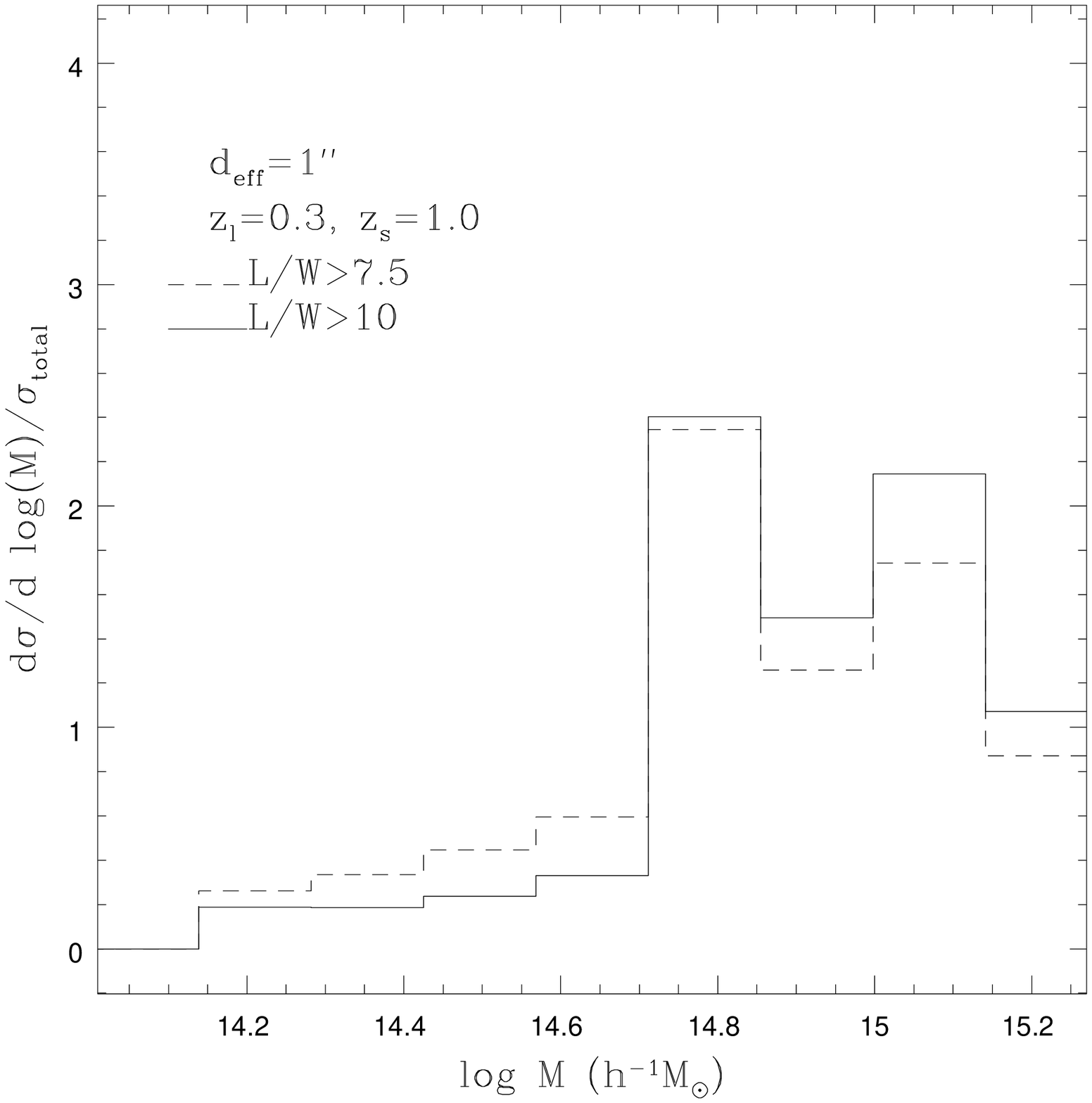} \\
\caption{The differential cross-section as a function of the logarithm
of the cluster mass for giant arcs with $L/W>7.5$ (dashed line) and
$L/W>10$ (solid line). The total area under each curve is normalized to unity.
The clusters are at redshift 0.2 (left panel) and 0.3 (right panel)
and the sources are at redshift 1.0.}
\label{fig:mlens}
\end{figure}

Fig. \ref{fig:mlens} shows how the cross-section of giant arcs is contributed
by clusters with different masses at redshifts 0.3 and 0.2. The background
source population is at redshift 1.0. For $\zl=0.3$ (right panel), the distribution is peaked around
$\sim 8\times 10^{14}h^{-1} M_\odot$, and it rapidly declines
below $\sim 2\times 10^{14} h^{-1} M_\odot$ as the much faster decrease
in cross-sections dominates over the rising numbers of low-mass clusters. The
cross-section also declines above $M\sim 2\times 10^{15} h^{-1} M_\odot$
due the exponential drop in the number of very massive clusters.
The left panel is for $\zl=0.2$, which can be directly compared with Fig. 4 in
Dalal et al. (2004) -- the two distributions are roughly consistent with
each other\footnote{Their axis labels in Fig. 4 should be $M_\odot$, not
$h^{-1} M_\odot$ (Dalal 2005, private communication).}.

\begin{figure}
\epsscale{0.8} \plotone{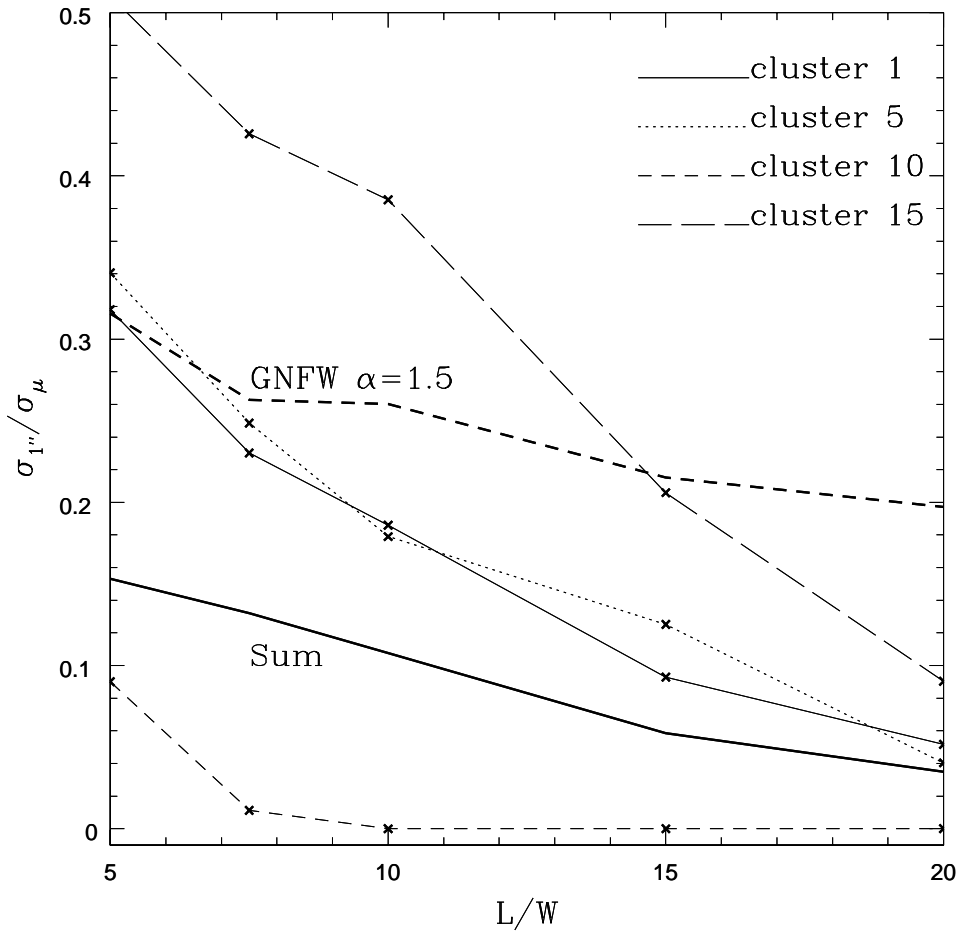} \caption{
The ratio of the cross-sections $\sigmasim$ and $\sigma_\mu$ as a
function of $L/W$
 for clusters at redshift 0.3 and background sources at redshift 1.0.
$\sigmasim$ is the cross-section obtained using ellipse-fitting, while
$\sigma_\mu$ is calculated assuming  that
the $L/W$ ratio is equal to the magnification.
The thick line is the cross-section weighted average for the 200 most massive clusters in our
simulations. The thick dashed line is for our generalized NFW model (see
eq. \ref{eq:gnfw}). The four thin lines are for the first, fifth, 
tenth and fifteenth most massive clusters (projected along the
$x$-direction) with
$M=1.70\times 10^{15}$, $7.58\times 10^{14}$, $6.55 \times 10^{14}$,
$6.02 \times 10^{14} h^{-1} M_\odot$ respectively.
}

\label{fig:sim2mu}
\end{figure}

The ratio of $\sigmasim$ and $\sigma_\mu$ is shown in Fig. \ref{fig:sim2mu}
as a function of the length-to-width ratio ($L/W$) for clusters
at redshift 0.3 and background sources at redshift 1.0.
The thick line is the cross-section weighted average for the 200 most
massive clusters while the thick dashed line shows 
the result for the generalized NFW model discussed above.
The four thin lines are the results for four individual clusters,
which are the first, fifth, 
tenth and fifteenth most massive clusters with 
$M=1.70 \times 10^{15}, 7.58\times 10^{14}, 6.55 \times 10^{14}$, and
$6.02 \times 10^{14} h^{-1} M_\odot$ respectively.

Recall that $\sigmasim$ is the cross-section calculated
using ellipse-fitting, while $\sigma_\mu$ is obtained
assuming that the $L/W$ ratio is equal to the magnification. 

Clearly, the assumption
that $L/W=|\mu|$ leads to an over-estimate of the cross-sections by a
factor of 7 and 10 for $L/W\geq7.5$ and $L/W\geq10$ respectively. 

\begin{figure}
\epsscale{0.8} \plotone{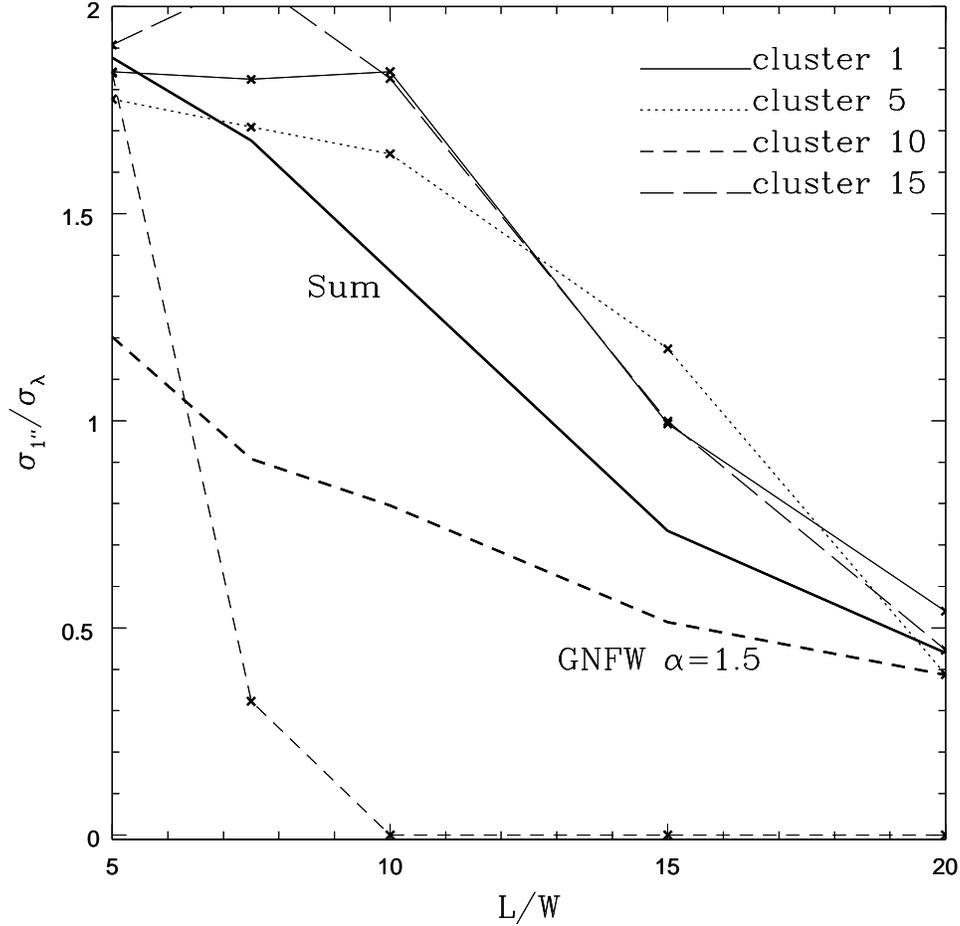} \caption{ 
The ratio of the two cross-sections, $\sigmasim$ and $\sigma_\lambda$, 
for clusters at redshift 0.3 and background sources at redshift 1.0.
For the two cross-sections,
the $L/W$ ratio is obtained through ellipse-fitting and under the assumption
that $L/W=|\lambda_1/\lambda_2|$. The thick line is the 
cross-section weighted average for the 200 most massive clusters in our
simulations while the thick 
dashed line is for our generalized NFW model (see eq. \ref{eq:gnfw}).
 The other four thin lines are for the first, fifth, 
tenth and fifteenth most massive clusters (projected along the
$x$-direction) with
$M=1.70\times 10^{15}$, $7.58\times 10^{14}$, $6.55 \times 10^{14}$,
$6.02 \times 10^{14} h^{-1} M_\odot$
 respectively.}

\label{fig:sim2lambda}
\end{figure}

The ratio of $\sigmasim$ and $\sigma_\lambda$ is shown in Fig. \ref{fig:sim2lambda},
where 
$\sigma_\lambda$ is again the cross-section calculated assuming $L/W$ to
be equal to the ratio of the two eigenvalues. $\sigmasim/\sigma_\lambda$ is
in the range of 0.5-2 when $L/W$ is in the range of 5-20. Thus,
$\sigma_\lambda$ offers a better approximation for the arc
cross-section than $\sigma_\mu$.
Notice that the more massive
a cluster is or the larger the caustics are, the higher 
$\sigmasim/\sigma_\mu$ and
$\sigmasim/\sigma_\lambda$ become. This can be understood as follows:
as the caustics become larger, the effect of finite
source size smoothing decreases. Thus,
$\sigmasim$ increases and so do its ratios with $\sigma_\mu$ and $\sigma_\lambda$.

\subsection{Optical depths}

To evaluate the optical depth, we first calculate the
 average cross-section per unit comoving volume:
\beq
\overline\sigma(\zl, \zs) = {\sum\sigma_i(\zl,\zs) \over V},
\eeq
where $\sigma_i(\zl, \zs)$ is the average cross-section of the 
three projections of the $i$-th 
cluster at redshift $\zl$, $\zs$ is the source redshift, and $V$ is
the comoving volume of the box adopted in our lensing simulations.
The optical depth can then be calculated as:
\beq \label{eq:tau}
\tau(\zs) = {{1\over{4\pi{\Ds}^2}} {\int_0}^{\zs}\, dz \, \overline\sigma(z,\zs)
(1+z)^3\, {dV_{\rm p}(z)\over dz}},
\eeq
where $\Ds$ is the angular diameter distance to the source plane, and
 $dV_{\rm p}(z)$ is 
the proper volume of a spherical shell with redshift from $z$ to $z+dz$.

Fig. \ref{fig:tau} shows the optical depth as a function of the source
redshift, $\zs$. Our results can be well fitted by the simple curve
\beq
\tau_{1''} = 2.25 \times 10^{-6}{1 \over 1+(\zs/3.14)^{-3.42}}
\eeq
within 20\%.

Fig. \ref{fig:dtaudz} shows the differential optical depth as a function
of the cluster redshift. For a source at redshift 1, 
there is little contribution to the optical depth for clusters
beyond redshift 0.7, because the critical surface density increases when the
lens is close to the source, and very few clusters are super-critical.
This trend is in excellent agreement with B98 (see their Fig. 1) and Dalal et al. (2004, 
see their Fig. 10). For a source at redshift 1, the optimal lensing
redshift is around 0.3, which is why we have chosen this redshift to illustrate many of our
results for $\zs=1$. For a source at redshift 2, the peak
shifts to about 0.5. For $\zs>3$, the optimal redshift is near
0.6-0.7, and the distributions also become noticeably
broader, with a full-width-at-half-maximum of about 0.9. This may have 
particular relevance to Gladders et al. (2003) who emphasized the lack
of lensing clusters with $\zl<0.64$ in their sample. There are likely 
selection biases in the surveys, but we notice that their four 
(approximate) source redshifts range from 1.7 to 4.9. If such source redshifts
are typical for their background galaxies, then the large cluster lens
redshift is fully consistent with our predictions. 
 As can be seen from Fig. \ref{fig:dtaudz}, for sources above
redshift 3, most lensing cross-sections are due to lenses below redshift 2.4, and there
are wiggles in the optical depth curves. The wiggles around $z_{\rm l}\sim
0.5$ are due to numerical effects as the interval of the lens redshift
(where particle positions are dumped from numerical simulations)
is large and the numerical integration is not exact. The
prominent wiggle around $\zl=1.5$ is, however, real. It arises due to significant merger
events in our simulations. For sources at high redshift, the
cross-sections are dominated by relatively few (most massive) clusters 
and merger events in these clusters have substantial impact on
the cross-section (Torri et al. 2004).

\subsection{Comparison with previous studies}
\label{sec:comparisons}

Our optical depth turns out to be
lower than three previous major studies (B98, W04 and
Dalal et al. 2004), by different amounts. We have traced 
these differences back primarily to the different values adopted for the
power-spectrum normalization ($\sigma_8$), how the $L/W$ ratio is measured, 
and whether the numerical simulations have sampled the cluster mass
function at the high-mass tail sufficiently. The
source size and ellipticity and the matter distribution
around clusters also have modest effects on the optical depth. Below we
discuss issues in detail.

\subsubsection{Comparison with B98}

In this study, we have followed closely the methodology of B98.
Nevertheless, our optical depth ($3.6\times 10^{-8}$)
is roughly $10\%$ of the value found by
B98 ($3.3 \times 10^{-7}$) for a source population at redshift 1.

Our lower optical depth is mainly due to the lower
$\sigma_8$ parameter adopted in the current study. In B98, the optical depth
was obtained as an average over two sets of simulations, one
with $\sigma_8=1.12$ and the other with $\sigma_8=0.9$.
However, the optical depth
is dominated by the former, as the clusters in the high $\sigma_8$ 
simulation form earlier and are also more centrally concentrated --
both increase the arc cross-section considerably, leading to a higher optical depth than in
the current work. Fig. \ref{fig:massFunc} shows
the mass function of halos at redshift 0.3 predicted by the extended
Press-Schechter formalism (e.g. Sheth \& Tormen 2002 and references therein; 
Press \& Schechter 1976). The cluster abundance
for $\sigma_8=1.12$ is a factor of $\sim 3.5$ higher
at $M\sim 10^{15} h^{-1} M_\odot$ than that  for $\sigma_8=0.9$. Even
the moderately larger $\sigma_8$ adopted by W04 (0.95) 
leads to a 50\%  increase in the abundance of clusters at 
at $M\sim 10^{15} h^{-1} M_\odot$ than that  for $\sigma_8=0.9$, which
can increase their optical depth  relative to ours (see the end of \S\ref{sec:comparisonsW04}).

\subsubsection{Comparison with W04}
\label{sec:comparisonsW04}

The method used by W04 is quite different from ours. They
adopted a multiple-lens plane approach, modeling the universe with a three-dimensional
matter distribution. In our study, we assume that the formation
of giant arcs is dominated by individual clusters, and
the large-scale structure along the line-of-sight is not important.
As mentioned above, another key difference between our study and W04
is how the $L/W$ ratio is measured: W04 assume the $L/W$ ratio to equal 
the magnification, while we measure the $L/W$ ratio through ellipse-fitting.
W04 effectively assumed infinitesimally small circular
sources, while we assume the sources to be elliptical and to have a finite size.
We will examine below how these different assumptions affect our predictions.

\begin{figure}
\epsscale{0.8} \plotone{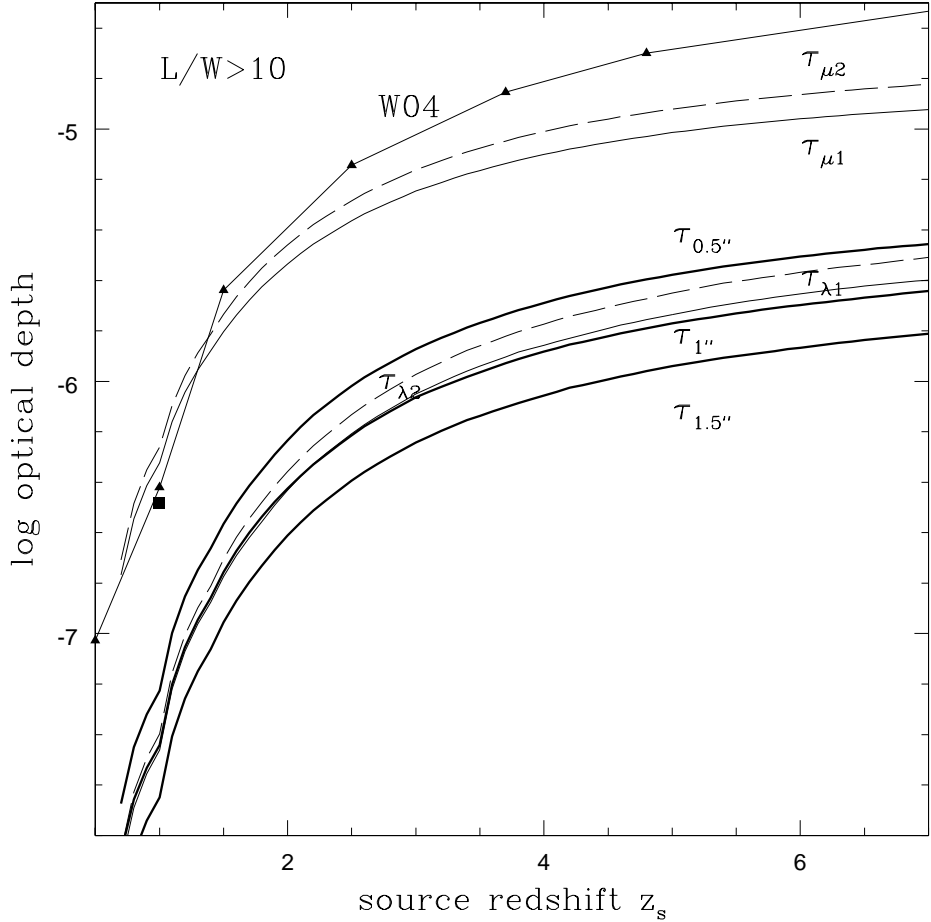} \caption{The optical depth as a
function of the source redshift for
giant arcs with $L/W>10$.  $\tau_{0.5''}$, $\tausim$, $\tau_{1.5''}$
are the optical depths calculated using ellipse-fitting assuming
an effective source diameter of $0.5\arcsec$, $1\arcsec$, and $1.5\arcsec$ respectively
(thick solid lines).
The thin solid curves labeled $\tau_{\mu1}$ and
$\tau_{\lambda1}$ are those calculated approximating $L/W$ by
the magnification and by the ratio of the two eigenvalues,
$|\lambda_1/\lambda_2|$, respectively. For these curves, the lensing potential
includes all the matter distribution
within a cube of $2\rv$. The curves labeled $\tau_{\mu2}$ and
$\tau_{\lambda2}$ are for the cases where we include all
the particles within a side length of $4\rv$ and a projection depth of
20$h^{-1}\,\mpc$. The triangles are the results of Wambsganss et
al. (2004) and the black square is that of B98.}
\label{fig:tau}
\end{figure}

\begin{figure}
\epsscale{0.8} 
\plotone{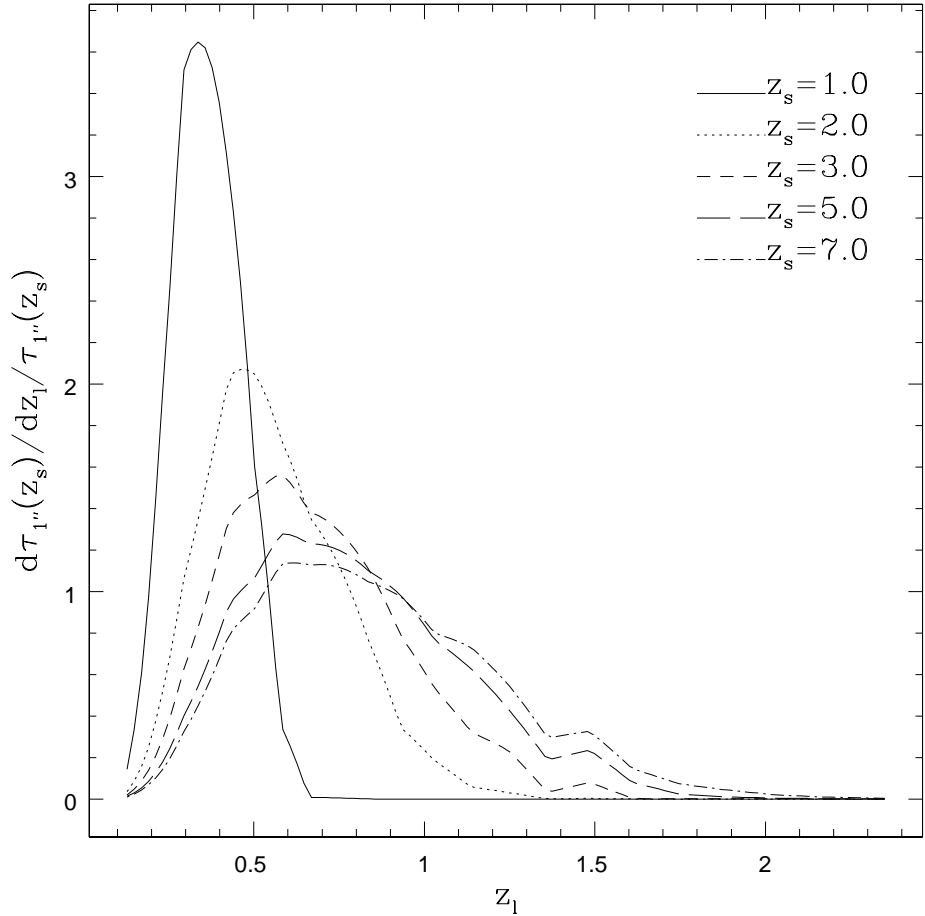} 
\caption{The differential lensing probability as a function of the lens
redshift. Curves are shown for five source redshifts, $\zs=1, 2, 3, 5,$
and $7$.
The area under each curve is normalized to unity. The wiggles
around $\zl=1.5$ for  sources at $z>3$ are due to cluster mergers
(see \S3.2).
}
\label{fig:dtaudz}
\end{figure}

If we adopt the same assumption as W04 (i.e., $L/W=|\mu|$), then our results are within
25\% of theirs (shown as triangles in Fig. \ref{fig:tau}) for a source redshift of $\zs=1$,
and within $50\%$ for $\zs$ out to $\sim 4.5$. 

We also qualitatively confirm the important result of W04 
that the optical depth is highly sensitive to the source redshift. Indeed, the optical
depth for a source at $\zs=1.5$ is a factor of 5 higher than
for a source at $\zs=1$. However,  there are some quantitative differences, namely,
their optical depth increases somewhat faster than ours
as a function of the source redshift; a similar discrepancy with W04
was found by Dalal et al. (2004). We return to this point at the end of
this subsection.

As we have discussed earlier in \S\ref{sec:caustics}, the
magnification is not a reliable estimator of the $L/W$ ratio. Instead,
 ellipse-fitting of simulated images
is needed to determine the $L/W$ value accurately. With this approach,
the optical depth is reduced by a factor of $\sim 10$
for a source at redshift unity
(see the thick solid curve labeled as $\tausim$ in Fig. \ref{fig:tau}).
The smaller value is a direct result of the smaller
cross-sections as we have shown in Figs.
\ref{fig:sim2mu}-\ref{fig:sim2lambda}.

In W04, the sources are assumed to be
infinitesimally small circular sources, while in our study the sources
are uniformly distributed in ellipticity from 0 to 0.5, and they
occupy an area equal to a circular source with a diameter of $1\arcsec$.

The finite source size effect may be important, particularly for
low-mass clusters whose caustics are small.  
Ferguson et al. (2004) recently used the HST to study the size evolution
of galaxies as a function of redshift. They found that for galaxies around redshift
1.4, the half-light diameters range between $0.4\arcsec$ to $2.2\arcsec$ with a peak around $1.4\arcsec$.
The size becomes smaller as the source redshift increases. The distributions are similar
for sources above redshift 2.3, with a peak around $0.5\arcsec$
and range from $0.2\arcsec$ to $2\arcsec$. Our adopted size falls within
the range for galaxies from redshift 1.4 to 5. As the redshift
of giant arcs spans a wide range (from 0.4 to 5.6, see
Table 1 in W04), it is important to examine
how our results change if we adopt difference source sizes.
Fig. \ref{fig:sourceSize} shows the cumulative cross-section as a function of cluster mass
for four source diameters, from $0.25$, 0.5, 1.0 and 1.5 arc seconds respectively.
In this exercise, we have taken the source population to be at redshift 1.0, and the
clusters are at redshift 0.3, i.e. approximately at the optimal cluster lensing
redshift for $\zs=1$ (see Fig. \ref{fig:dtaudz}). The cumulative
cross-sections are calculated down to $M\sim 2.2\times
10^{14}h^{-1}M_\odot$, below which clusters do not contribute
substantial cross-sections (see Fig. \ref{fig:mlens}).
For giant arcs with $L/W>10$, the cross-section
increases by about 50\% when the source diameter changes from $1\arcsec$ to
$0.25\arcsec$. For giant arcs with $L/W>7.5$ and $L/W>5$, as expected,
the finite source effects are even more modest. The optical depths
for $\deff=0.5\arcsec, 1.5\arcsec$ are shown as functions of the source
redshift in Fig. \ref{fig:tau}. The optical depth increases by about 50\% 
as $\deff$ decreases from $1\arcsec$ to $0.5\arcsec$. 
The finite source size effect appears modest, insufficient to
explain our discrepancy with W04. Notice that
$\tau_{1''}$ agrees well with $\tau_\lambda$ for nearly all source redshifts.
Furthermore, notice that as the source size decreases from $1\arcsec$ to $0.5\arcsec$,
the optical depth does not follow $\tau_\lambda$ any more closely, as one
might expect. This arises because, following W04, we have used only
the brightest of multiple images for
$\sigma_\lambda$. In contrast, when calculating $\sigma_{0.5''}$,
$\sigma_{1''}$ and so on, the cross-section has been
multiplied by the number of giant arcs that satisfy the selection criteria
(a similar procedure was used by B98). So in general, even
when the source size decreases to zero, these two cross-sections do not
overlap with each other.

Another difference between our study and W04 is that we use
elliptical sources while W04 use (infinitesimally small) circular sources.
The ellipticity distribution we adopt (uniform between 0 to 0.5) 
is in good agreement with the study of Ferguson et al. (2004)
who found a flat distribution between 0.1 to 0.6 (with a small
drop in the number of galaxies with zero ellipticity). Nevertheless,
we want to check how the ellipticity changes the results.
Fig. \ref{fig:ellipticity} shows the ratio of the cumulative
cross-sections as a function of cluster mass for
elliptical sources and circular sources for giant arcs with $L/W$ larger
than 5, 7.5 and 10. Again 
all the clusters are at redshift 0.3 and the sources are placed at redshift 1.
In all cases, the ellipticity
increases the cross-section by 20\% to 55\%, implying
that if we adopted circular sources, the discrepancy with W04 will
become slightly worse. Notice 
that the rate of increase depends both on the source size and the $L/W$ ratio,
but the dependence is not monotonic due to the competing effects of finite
source size and ellipticity, as was also found by Oguri (2002). In summary,
the finite source
size and ellipticity of sources both have only modest effects on the optical depth
of giant arcs, insufficient to explain the
difference between our study and W04.

As we mentioned above, we reproduce the optical depth of W04
if we approximate the $L/W$ by magnification. However, rigorous 
 ellipse-fitting of simulated images reduce the optical depth by a
factor of $\sim$ 
10. This means only $\sim 10\%$ or so of the images with $|\mu|>10$ form giant arcs with
$L/W>10$. The question thus naturally arises: why do the rest 90\% of images fail to
make giant arcs $L/W>10$? First, it is well-known that for isothermal spheres ($\rho\propto r^{-2}$),
the $L/W$ ratio is identical to the magnification. But clusters are
not isothermal spheres. As we have illustrated
using a generalized NFW profile (a better approximation for clusters) in
\S\ref{sec:caustics}, the magnification is no longer equal to the $L/W$ ratio.

Second, the top right panel in Fig. \ref{fig:4p} clearly shows that, for
this cluster, most giant arcs form at the position of the
cusps while few giant arcs are formed near fold sections of caustics (which
contribute most of the high magnification images)
in the source plane. Images 
with high magnification and low distortion must thus be along caustics but away 
from cusps. This is also expected from catastrophe theory, 
because images with high magnifications and low distortions are formed
near the so-called 
``lips'' or ``beak-to-beak'' caustics (cf.~Schneider et al.~1992).

However, the question still remains: what are the image configurations of the rest of the highly
magnified images? Fig. \ref{fig:fraction} provides the answer. It shows that roughly
50\% and 80\% of the highly magnified images with magnification above 10
form arcs with $L/W>5$ and $L/W>3$ respectively; here the magnification
is determined by the ratio of the image area and the source area. The
remaining 20\% or so images are highly magnified but (largely) undistorted (HMUs). 
Williams \& Lewis (1998) studied such images for cored isothermal
and NFW density profiles. They found that for isothermal cored clusters,
the ratio between HMUs and giant arcs with $L/W>10$ is of the order
of unity. For NFW profiles, the ratio is only about 30\%. Our corresponding ratio
is about $\sim 2$  and lower for smaller sources.
Notice that there are comparable number of images
with $L/W>10$ and $|\mu|<10$. This will further decrease the relative number
of HMUs relative to the observed number of giant arcs with $L/W>10$.
We do not regard this as a serious discrepancy, as 
Williams \& Lewis (1998) used spherically symmetric clusters as lenses.
Cluster lenses forming large arcs are preferentially merging and highly irregular, so the
effect of shear due to substructures may be much more important than that
in their studies, which can plausibly explain the differences in the ratio of giant 
arcs and HMUs.

A related question concerns the width of giant arcs, which provides valuable
information about the inner density profile of clusters. To see the
general trend, we use a spherically symmetric toy model for clusters,
where the density is a power-law as a function of radius,
$\rho \propto r^{-\beta}$. For an isothermal sphere, $\beta=2$, and the
inner slope  of an NFW profile is $\beta=1$.
For an infinitesimally small circular source, when $\mu \gg 1$,
the width of a tangential arc is equal to $1/(\beta-1)$ times the
diameter of the (circular) source. For an isothermal sphere, lensing
conserves the source width. A point lens corresponds to the limit $\beta
\rightarrow 3$, in
this case the width is equal to one half of the source diameter. In reality,
clusters are not spherically symmetric, nevertheless, the width and
length of giant arcs carry valuable information about the central density profiles in
clusters (Hammer 1991; Miralda-Escud\'e 1993).
Fig. \ref{fig:widthLW} shows the distribution of widths of giant arcs with $L/W>7.5$. For
$\deff=1\arcsec$, the distribution peaks around $1.4\arcsec$, and
rapidly drops off at $2\arcsec$. For
$\deff=0.5\arcsec$, the distribution peaks around $0.7\arcsec$ and
drops off rapidly around $1\arcsec$, except for the tenth most massive cluster which
seems to have a much broader distribution. This is because this cluster has
the smallest caustics, and the finite source size is most important
in this case. The thick solid histogram shows the observed
width distribution in Sand et al. (2005) for giant arc candidates with $L/W>7.5$ (in our
definition). This comparison does not yet fully account for the
(unknown) redshift distribution of background sources and other
potential biases. Nevertheless, it appears very interesting
that the size distribution is in better
agreement with an effective source diameter of $\deff=0.5\arcsec$,
although the tail indicates that larger sources are also needed.

W04 emphasized the effect of matter along the line of sight on the
arc statistics. We are unable to model the universe using multiple
lens planes,
as realistically performed by W04. We can, however, 
investigate the effect of matter distribution (e.g., filaments)
surrounding the clusters. To do this, we increase the
projection depth to $20h^{-1}$ Mpc (comoving) and the lens area to a
square with a side-length of $4\rv$. We write the new optical depths as
$\tau_{\lambda2}$ and $\tau_{\mu2}$, in contrast to the values obtained using a cube
of side length $2\rv$, denoted by 
$\tau_{\lambda1}$, $\tau_{\mu1}$. Fig. \ref{fig:tau} shows these optical
depths as functions of the source redshift. Clearly the matter
distribution around the clusters increases the optical depth but the effect
appears to be small ($\sim 15\%$).
If the enhancement due to the large-scale structure along the line-of-sight
is more important (W04), then the optical depth
at higher source redshift will be affected more, which may partially
explain why our optical depth increases somewhat more slowly as a function of
source redshift than W04, although Hennawi, Dalal \& Ostriker (2005) showed that 
multi-plane lensing changes the cross-section only slightly compared with the single-plane
lensing (see their Fig. 2).
In addition W04 used a power-spectrum normalization
$\sigma_8=0.95$, slightly higher than our value ($\sigma_8=0.9$). This
will also lead to a slightly higher optical depth because in their
simulation, the clusters will form somewhat earlier, which
will in turn preferentially boost the
optical depth for a source at higher redshift as their optimal lens
redshift is higher (see Fig. \ref{fig:dtaudz}).

\subsubsection{Comparison with Dalal et al. (2004)}

Most of our conclusions agree with those of Dalal et al. (2004), who
used numerically-simulated clusters from the GIF collaboration (Kauffmann et al. 1999).
Their mass and spatial resolutions are similar to ours. However,
their overall optical depths are larger than ours by a factor of
6 for $\zs=1.0$. The difference is partly due to the different definitions of 
the $L/W$ ratio. Dalal et al. fit the giant arcs
by a rectangle, rather than an ellipse as in our case.
Thus for the same giant arc, the $L/W$ ratio in
Dalal et al. (2004) is equal to $4/\pi$ times our $L/W$
ratio. As a result, their optical depth for giant arcs with $L/W\geq10$ 
should be compared with our value for $L/W \geq 7.5$.
For example, for $\zs=1, 1.5$ and 2, our optical depths for giant arcs
with $L/W\geq7.5$ (in our definition) are
$1.0\times 10^{-7}, 4.8\times 10^{-7}$ and
$1.0 \times 10^{-6}$ respectively. 
These should be compared with their corresponding values
of $2.5\times 10^{-7}$, $7\times 10^{-7}$, and
$1.4\times 10^{-6}$. Our value is about 70\% of their value
for a source at redshift 1.5 and 2. But the discrepancy becomes larger
for a source at redshift 1.  At such low redshift, the cross-section
is increasingly dominated by the few most massive clusters as
the critical surface density increases,
so the cosmic variance becomes important, particularly for
the GIF simulations which have a simulation volume 9.6 times smaller than ours.
A detailed comparison 
between our numerical methods indicate that our cross-sections agree
within $25\%$ for the same numerical cluster 
(with the same projection)
selected from the GIF simulations (Dalal 2005, private
communication)\footnote{We developed two independent numerical 
codes. The cross-sections
from these two methods for the cluster 
shown in Fig. \ref{fig:4p} agree within 10\%-30\%.}. We believe that
the remaining difference is due to the different mass functions
in our simulations and the fact that Dalal et al. (2004) used
more than three projections to calculate the cross-sections
for the more massive clusters. Fig. \ref{fig:massFunc} shows that
the GIF mass function at large mass appears slightly higher than ours.
This will increase their optical depths for giant arcs and likely explain
the remaining difference.

\section{Discussion}

In this paper we used high-resolution numerically-simulated clusters to study the
rate of giant arcs in clusters of galaxies. The methodology we used is 
similar to B98, but the number of clusters in our simulations
is about a factor of $\sim 8-10$ higher than previous studies (e.g., B98 and
Dalal et al. 2004), so it should provide better statistics for the most
massive clusters. We calculated the cross-section and optical depth as a function of the
source redshift using ellipse-fitting and compared our results with those 
obtained with two approximate measures of the length-to-width ratio ($L/W$).
We also found that the effects of source size and ellipticity on the
optical depth are
modest. Furthermore, the matter distribution around clusters increases
the lensing cross-section only slightly, by $\sim 15\%$.

Detailed comparisons of our results were made with 
three major previous studies, B98, W04 and Dalal et al. (2004).

\begin{figure}
\epsscale{0.8}
\plotone{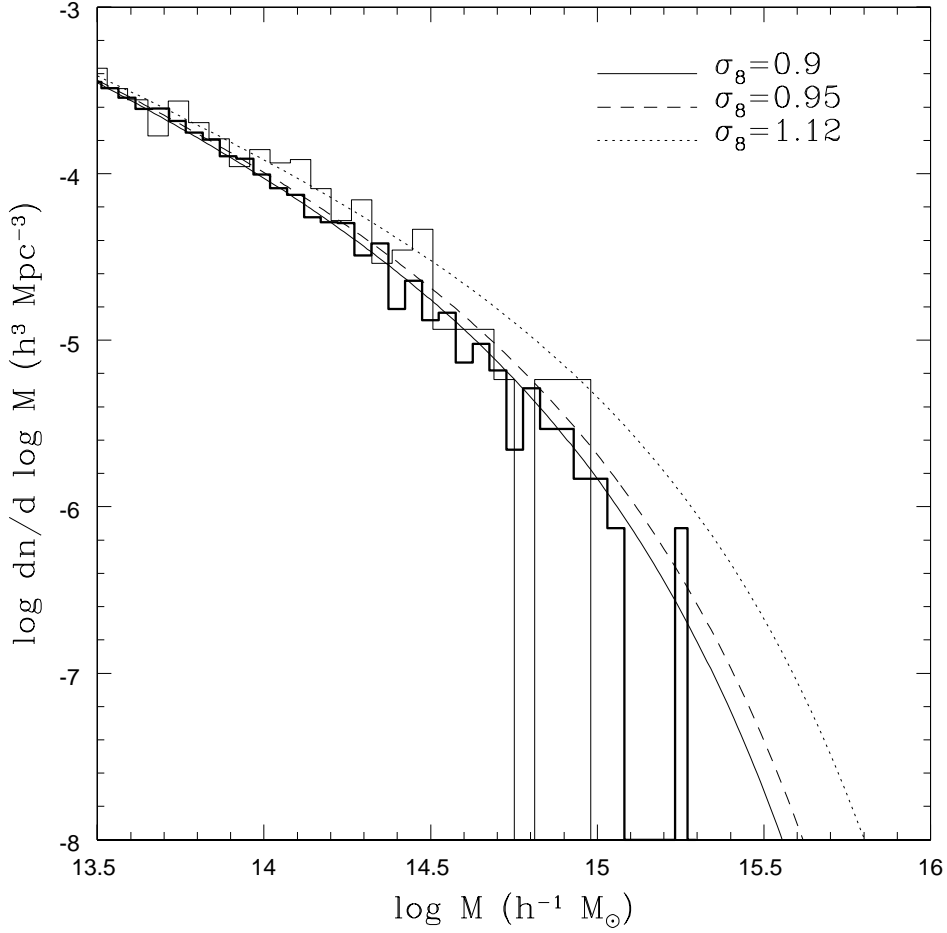}
\caption{ 
The number density of halos as a function of mass at redshift 0.3 as
predicted by the extended Press-Schechter formalism modified by 
Sheth \& Tormen (2002) in the $\Lambda$CDM model with different power-spectrum
normalizations, $\sigma_8$.
The solid line is for $\sigma_8=0.9$, while the dashed and dotted lines
are for $\sigma_8=0.95$ and $\sigma_8=1.12$ respectively.
Notice the large difference in the
abundance of clusters of galaxies at large masses. The bold histogram shows
the mass function
for our simulation while the thin histogram
shows that for the GIF simulation, both for $\sigma_8=0.9$.
}
\label{fig:massFunc}
\end{figure}

\begin{figure}
\epsscale{0.8}
\plotone{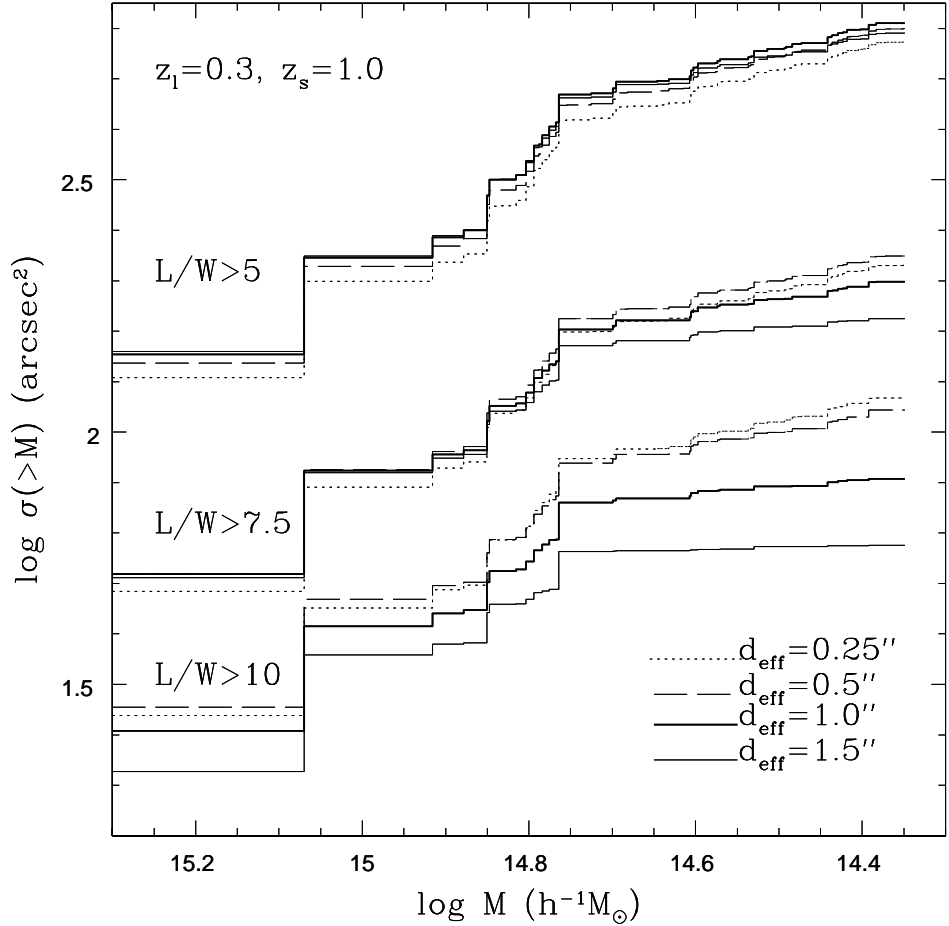}
\caption{The cumulative cross-sections, $\sigma(>M)$, as a function of
cluster mass for giant arcs with $L/W$ larger than 5 (top), 7.5
(middle), and 10 (bottom) respectively. Four curves are shown
for each $L/W$ ratio for four effective source diameters of $\deff=0.25\arcsec,
0.5\arcsec, 1.0\arcsec$ and $1.5\arcsec$ respectively.
The clusters are at redshift 0.3 and sources at redshift 1.0.}
\label{fig:sourceSize}
\end{figure}

\begin{figure}
\epsscale{0.8} 
\plotone{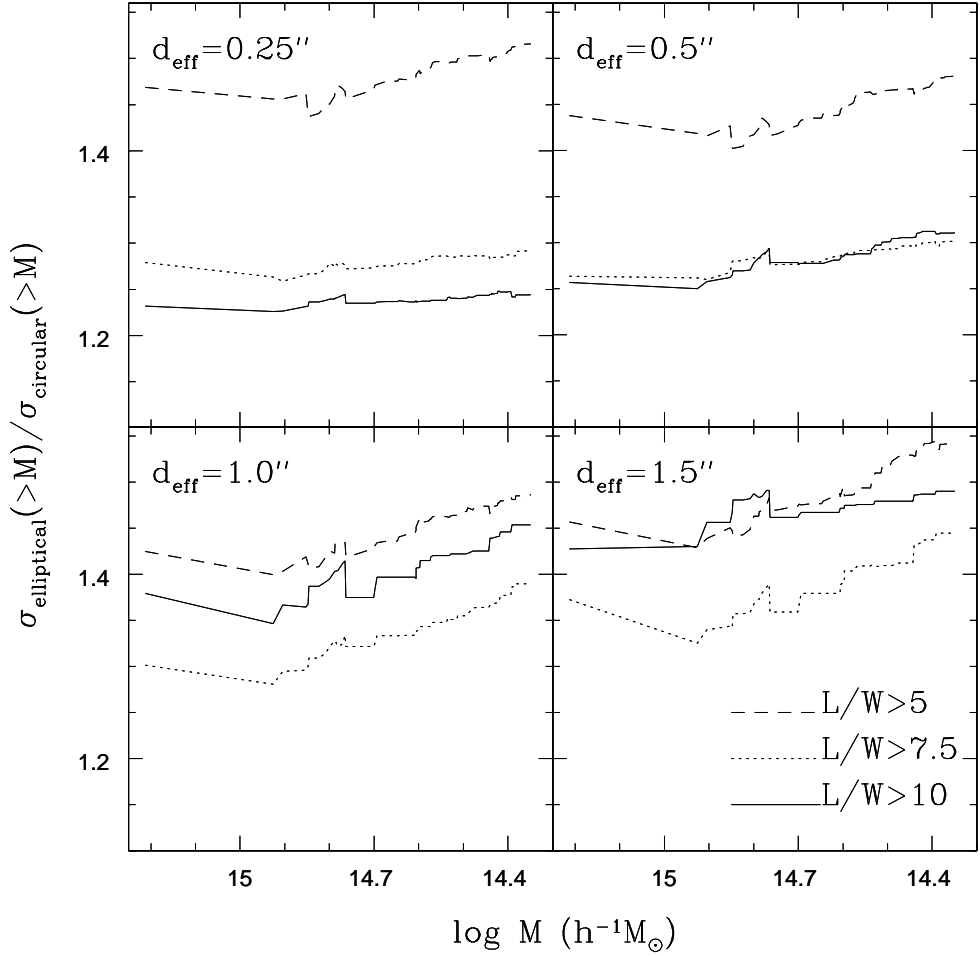}
\caption{The ratio of the cumulative cross-section,
$\sigma_{\rm elliptical}(>M)$, and $\sigma_{\rm circular}(>M)$,
where the sources
are modeled as elliptical and circular sources respectively. 
Results are shown for giant arcs with $L/W$ larger than 5 (dashed), 7.5
(dotted) and 10 (solid) and four
effective source diameters,  $\deff=0.25\arcsec, 0.5\arcsec, 1\arcsec$, and
$1.5\arcsec$.
The clusters are at redshift 0.3 and sources at redshift 1.0.}
\label{fig:ellipticity}
\end{figure}

\begin{figure}
\epsscale{0.8} 
\plotone{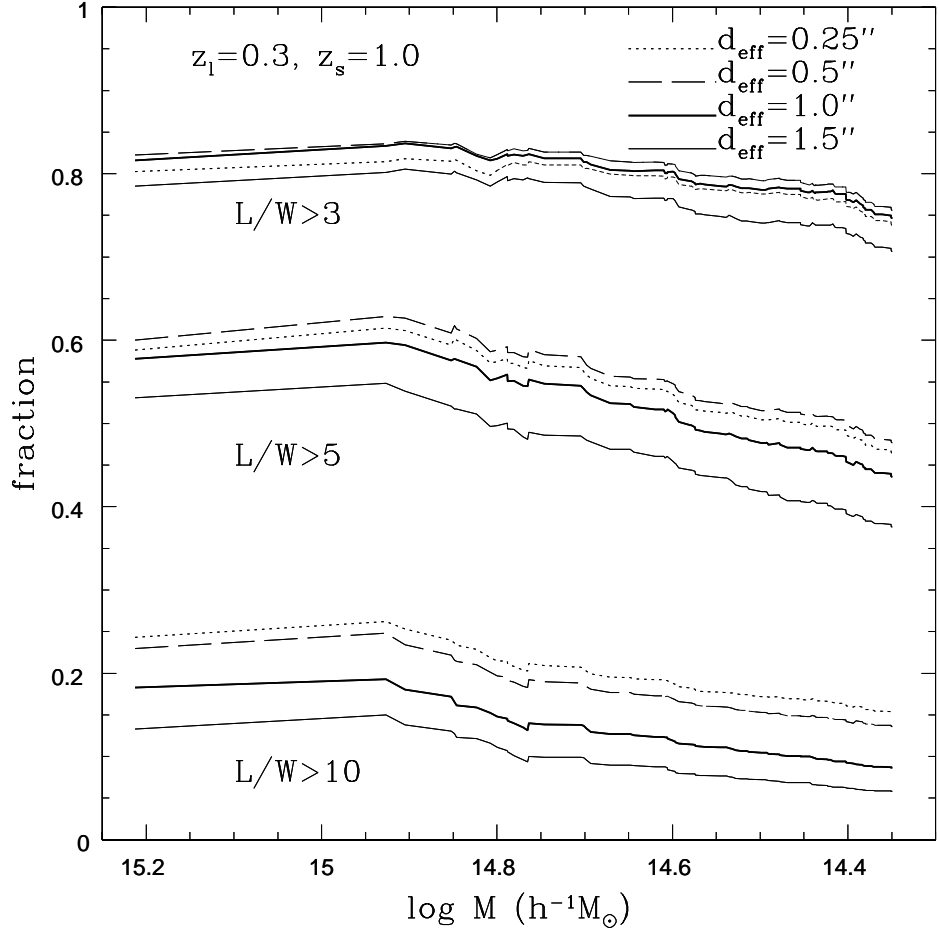}
\caption{The ratio of the cumulative cross-sections, 
$\sigma(>|\mu|, >L/W)/\sigma(>\mu)$, as a function
of the cluster mass. All the sources have magnifications exceeding 10.
Results are shown for arcs with $L/W$ larger than 3, 5 and 10 and four
effective source diameters,  $\deff=0.25\arcsec, 0.5\arcsec, 1\arcsec$, and
$1.5\arcsec$.
The clusters are at redshift 0.3 and sources at redshift 1.0.
}
\label{fig:fraction}
\end{figure}

Our optical depth is about 10\% of that found B98. The difference arises because
the two studies used two different $\sigma_8$ values -- B98 adopted a higher
$\sigma_8$ (1.12) than ours (0.9). The higher $\sigma_8$ value leads to
both a higher number density and more concentrated clusters
(see Fig. \ref{fig:massFunc}). Both effects increase their optical depth.

Our results qualitatively confirm one important
conclusion reached by W04, namely that the optical depth is sensitive to
the source redshift, but the rate of increase in our simulations is
somewhat slower than theirs. The large-scale structure, which was included in W04 but
ignored by us, will become more
important as the source redshift increases, but whether this effect can
explain the discrepancy in full is unclear (see also Dalal et
al. 2004). Perhaps more importantly, we find that
their assumption, $L/W=|\mu|$,  over-estimates the number of
giant arcs by a factor of $\sim 10$. 

Better agreements were found between our study and 
Dalal et al. (2004). Once we account for the difference in the arc
modeling (ellipses vs. rectangles), our optical depth
for giant arcs with $L/W \geq 7.5$ is about 70\% of that found by
Dalal et al. (2004) for a source at redshift 1.5 and 2.0. For a source
at redshift 1, our prediction is about 40\% of their value.
We believe our differences are due to cosmic variance --
the GIF simulation has a cosmic volume that is 9.6 times smaller
than ours, and their mass function appears to be
somewhat higher than ours (see Fig. \ref{fig:massFunc}), which may
explain their higher optical depth. 

Both W04 and Dalal et al. (2004) concluded that the predicted and observed
rates of giant arcs are consistent with each other. However, the
comparison study we have performed leaves this question open. 
It appears to us that our predicted rate may be a factor of a
few too low compared with the ``observed'' rate. However,
this conclusion is not firm as both
observations and theoretical predictions are uncertain. Observationally,
we need more transparent selection criteria
(see Dalal et al. 2004 for excellent discussions). Furthermore, 
the $L/W$ determination from ground-based telescopes may be
affected by seeing, leading to a likely under-estimate of the $L/W$ ratio
as the widths of many arcs are unresolved. 

The recent extensive search for giant arcs with HST images 
(Sand et al. 2005) is a significant step in the right
direction. Observers need to report not only the $L/W$ ratio but also
individual widths, which can provide important constraints on the inner 
density profile in clusters (see \S\ref{sec:comparisonsW04}). 
Our predicted width distribution appears to better match the 
observed distribution of Sand et al. (2005) when the effective source
diameter $0.5\arcsec$. However,

a reliable prediction
of the giant arcs also requires accurate information on the source
population, including their redshift, size, magnitude, surface brightness and ellipticity. 
Such information is of course a key objective for studying high-redshift
galaxies. As the information is currently lacking, the prediction for
the number of giant arcs has yet to enter the high-precision era.

Theoretically,  we need
higher resolution simulations in very large boxes so that
we can sample the cluster mass function and
resolve the internal structure of clusters simultaneously. Numerical
simulations with baryonic cooling and star formation will also be
needed to make more detailed comparisons with observational data.
While baryonic cooling
is not expected to be very efficient in clusters of galaxies (most
baryons in clusters are still in the hot phase seen as X-rays),
nevertheless, its effects may not be negligible, particularly for the
giant arcs at small radii (Dalal et al. 2004; see also Oguri 2003). 
In fact, the recent study by 
Puchwein et al. (2005) found that the effects of baryons
on giant arcs depend on the detailed implementation of viscosity, star
formation and feedback processes in simulations. If these effects can be
implemented in practice, hydro-dynamical cosmological
simulations also offer the possibility of
a more direct comparison with observations, at least with the
EMSS, as the X-ray luminosities of clusters of galaxies in
these simulations can be
predicted (with some uncertainty), and hence we can apply similar
selection criteria for clusters of galaxies as in observations
and examine the number of giant arcs in these clusters.

As many observations converge to the concordance cosmology, it will be
interesting to use lensing to constrain parameters such as
$\sigma_8$ independently in the $\Omega_{\rm m,0}=0.3, \Omega_{\Lambda, 0}=0.7$ flat
cosmology. As the cluster
mass function and internal structures both sensitively depend on $\sigma_8$, the arc statistics
should provide a stringent limit on $\sigma_8$, as already illustrated
by the difference between the current study and B98. This parameter is
still somewhat uncertain: some studies prefer values as high as 1.1,
while others prefer values as low as 0.7 (see Tegmark 2004 for a recent
overview of current results).
We plan to return to some of these issues in the future.

\begin{figure}
\epsscale{1.1} \plottwo{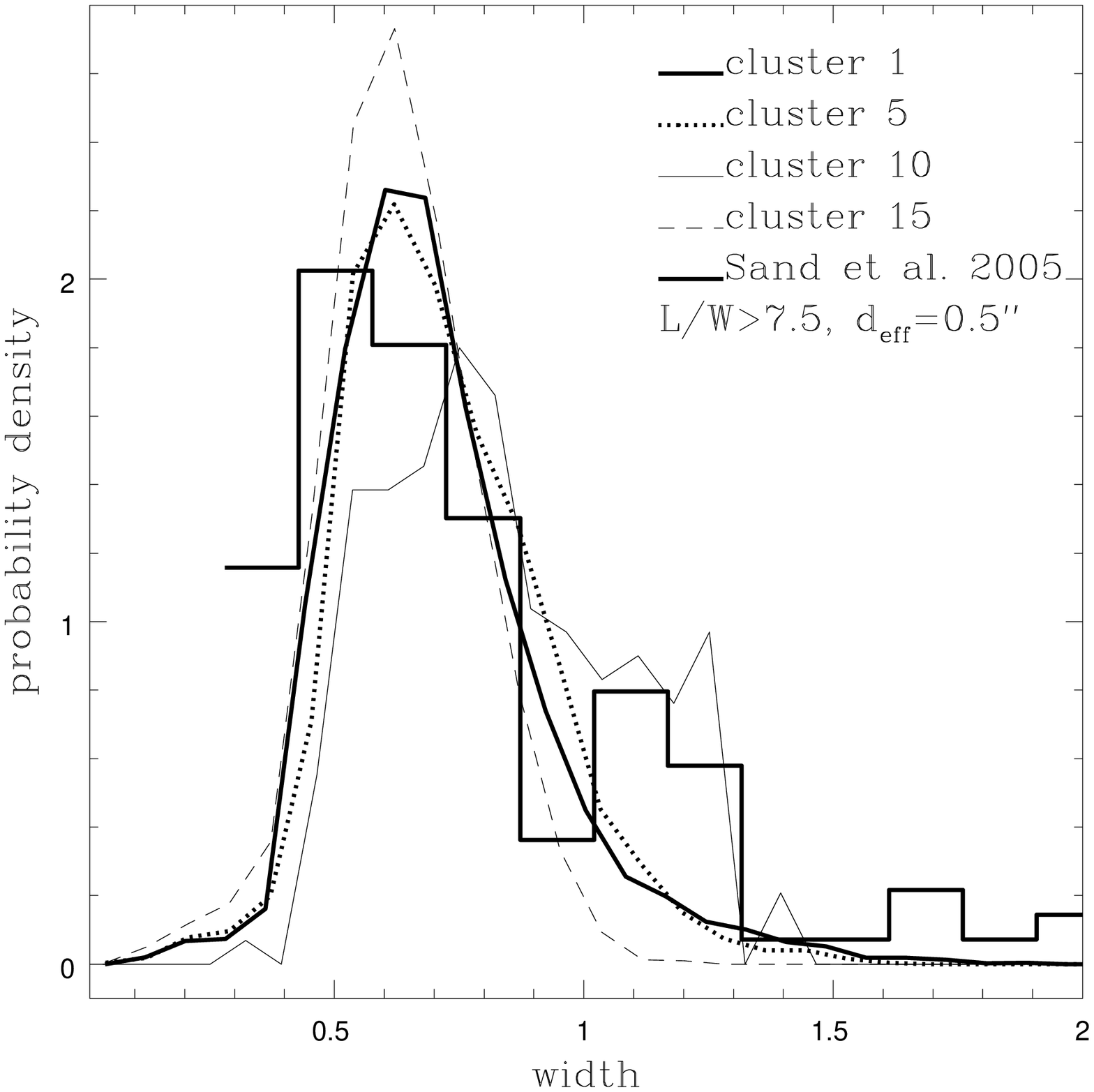}{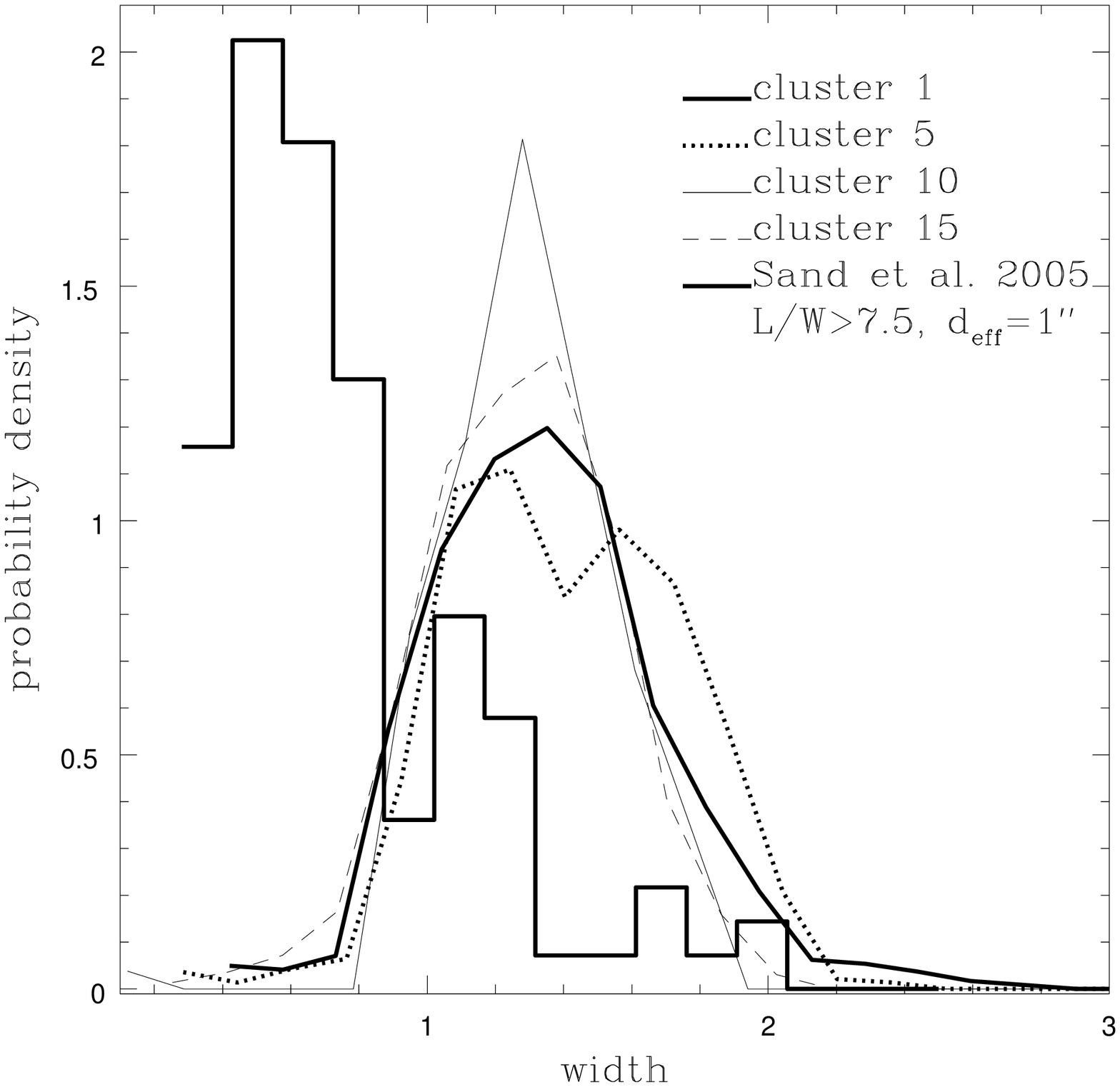}
\caption{The distribution of widths for giant arcs with $L/W>7.5$
for the four clusters (at redshift 0.3) shown in Figs. \ref{fig:sim2mu} and
\ref{fig:sim2lambda}. The sources are at redshift 1.0
and are assumed to have an effective source diameter of $0.5\arcsec$
(left panel) and $1.0\arcsec$ (right panel) respectively. The thick
solid histogram is for the data of Sand et al. (2005).
}
\label{fig:widthLW}
\end{figure}

\acknowledgments
We thank Drs. L. Gao, W. P. Lin and in particular N. Dalal for helpful
discussions. We are grateful to David Sand for the data plotted
in Fig. \ref{fig:widthLW}. We are also indebted to an anonymous referee 
for an insightful report that improved the paper. 
YPJ is supported in part by NKBRSF (G19990754) and by NSFC.
SM and LGL acknowledge the financial support of Chinese Academy
of Sciences and the European Community's Sixth Framework Marie 
Curie Research Training Network Programme, Contract No. 
MRTN-CT-2004-505183 ``ANGLES".

{}

\end{document}